\documentclass[showpacs,twocolumn,aps]{revtex4-1}
\usepackage[english]{babel}
\usepackage[utf8]{inputenc}
\usepackage{graphicx}
\usepackage{epstopdf}
\usepackage{amsmath,amssymb}
\usepackage{dcolumn}
\usepackage{subfigure}
\usepackage{amsmath,amsfonts,amsthm,bm}
\usepackage{mathtools}
\usepackage{float}

\begin{document}

\title{Measuring a QED cross section via a witness particle}

\author{Jonas B. Araujo $^a$}\email[]{jonas.araujo88@gmail.com}
\author{B. Hiller $^b$}\email[]{brigitte@fis.uc.pt}
\author{I. G. da Paz $^c$}\email[]{irismarpaz@ufpi.edu.br}
\author{Manoel M. Ferreira Jr. $^a$}\email[]{manojr.ufma@gmail.com}
\author{Marcos Sampaio$^d$}\email[]{marcos.sampaio@ufabc.edu.br}
\author{H. A. S. Costa $^c$}\email[]{helderfisica@gmail.com}

\affiliation{$^a$ Universidade Federal do Maranh\~ao, Centro de Ci\^{e}ncias Exatas e Tecnologia, 65080-040, S\~ao Lu\'{\i}s, MA, Brazil}
\affiliation{$^{b}$ CFisUC, Department of Physics, University of Coimbra, P-3004-516 Coimbra, Portugal}
\affiliation{$^c$ Universidade Federal do Piau\'{\i}, Departamento de F\'{\i}sica, 64049-550, Teresina, PI, Brazil}
\affiliation{$^{d}$ CCNH, Universidade Federal do ABC,  09210-580 , Santo Andr\'e - SP, Brazil}

\begin{abstract}
\noindent
We consider a QED scattering ($AB\rightarrow AB$), in which $B$ is initially entangled with a third particle ($C$) that does not participate directly in the scattering. The effect of the scattering over $C$'s final state is evaluated and we note coherence (off-diagonal) terms are created, which lead to non null values for $\langle \sigma_x\rangle$ and $\langle \sigma_y\rangle$ that are, in principle, measurable in a Stern-Gerlach apparatus. We chose a particular QED scattering ($e^+e^-\rightarrow\mu^+\mu^-$) and found that $\langle \sigma_x\rangle$ and $\langle \sigma_y\rangle$ are proportional to the total cross section ($\sigma_{\text{total}}$) of the $AB$ scattering, besides being maximal if $BC$'s initial state is taken as a Bell basis. Furthermore, we calculated the initial and final mutual informations $I_{AC}$ and $I_{BC}$, and noticed an increase (decrease) in $I_{AC}$ ($I_{BC}$), which indicates that, after $AB$ interact, the total amount of correlations (quantum $+$ classical) are distributed among the $3$ subsystems.
\end{abstract}

\maketitle
\section{Introduction}
Arguably the most intriguing feature of quantum mechanics, entanglement has been shown 
to be a fundamental phenomenon in nature. About thirty years after the posing of the EPR 
paradox \cite{EPR}, which rebuked entanglement based on causality and locality arguments,
 Bell provided a  test \cite{Bell}, which was later implemented experimentally by Aspect 
\textit{et al.}, using polarization-entangled photons emitted by a calcium source \cite{Aspect}.  
Loop-holes in the experimental tests have been successively removed; recently, violations on Bell's inequality were measured for spins separated by $1.3$km \cite{Clauser}, and for light from distant astronomic sources \cite{CosmicBell}.

Regarding the technological applicability, entanglement plays a central role in the long-sought quantum computers \cite{Horodecki}, quantum metrology, quantum optics and 
optomechanical systems \cite{Metrology,Gravity}. In high energy physics, entanglement has  recently received considerable attention, mainly concerning the production of entropy in scattering processes -- for a description of entanglement generation in nonrelativistic quantum mechanics, see Ref. \cite{Mishima}. In quantum field theory (QFT), it has been studied, for example: the variation in 
 entanglement entropy in a relativistic scattering involving scalar fields \cite{Seki} -- one-loop calculations were done in \cite{Faleiro}, and the entropy generation of fermions systems in
  QED processes \cite{Fan2017,Fan2018}, in which the authors studied the mutual 
  information between spin degrees of freedom and properties of the entropy variation under Lorentz transformations. An interesting application related to metabolic PET-imaging 
 (Positron-Emission-Tomograph) is found in Ref. \cite{Compton}, in which a method to 
 detect entanglement of photons from positronium decays is proposed. In other recent 
 works, it was shown that entanglement can be used to magnify the photon-photon scattering cross section \cite{photon} and to enhance possible Lorentz symmetry violation effects in Yb$^+$ atoms \cite{LVth,LVexp}. These are applications of what is known as relativistic quantum information.
  
 In relativistic scenarios, such as QFT processes, it is fundamental to define 
 Lorentz-invariant entanglement measures. It has been shown that, for bipartite fermion systems, the linear entropy of each particle, considering both its spin and momentum, is Lorentz-invariant \cite{Bertlman2010,Fan2017}. Entanglement in the spin-spin partition, 
 although its entropy is not Lorentz-invariant, has been shown to violate the Clauser-Horne-Shimony (CHSH) inequality in the relativistic regime \cite{Marta2017}. As for the momentum-momentum partition, the dynamics of entanglement in lowest order QED has been studied, for instance, in \cite{Lamata}. Another fundamental aspect is the connection between maximal entanglement and gauge symmetries in QFT, studied for example in \cite{Alba2017}. 

Entanglement also plays a role in inflationary models described by QFT in curved spacetimes. It has been shown that an expanding spacetime could create fermion pairs that are entangled in opposite momentum modes \cite{Fuentes2010} -- the effect of QED in this process has been recently assessed in \cite{Lucas2018}. In the free case, it was possible to read from the fermion's von Neumann entropy the parameters of the expansion of the universe. It is important to point out that in these models there are fundamental differences between the fermionic and bosonic cases \cite{Fuentes2006,Helder1}. More realistic features, such as decoherence, have also been studied in QFT in expanding spacetimes \cite{Gustavo2014}.

 In this work, we study a QED scattering ($AB\rightarrow AB$) in which $B$ is initially entangled with a witness particle ($C$). The purpose is to extract information about the scattering by observing particle $C$. The paper is organized as follows. in Section \ref{DefsAndReducedC}, definitions are made and the final reduced density matrix of particle $C$ is calculated; we find that coherence terms are generated and evaluate their effect on particle $C$'s spin measures in different directions. in Section \ref{Mutualinfo} we analyze the change in mutual information between particles $A$-$C$, and $B$-$C$ due to the scattering; the results are consistent with a distribution of correlations (quantum$+$classical) among the subsystems $A$, $B$ and $C$.  The conclusions and final remarks are done in Section \ref{Conclusions}.

\section{Scattering with a witness particle}\label{DefsAndReducedC}
\begin{figure}[h]
\begin{centering}
\includegraphics[scale=0.45]{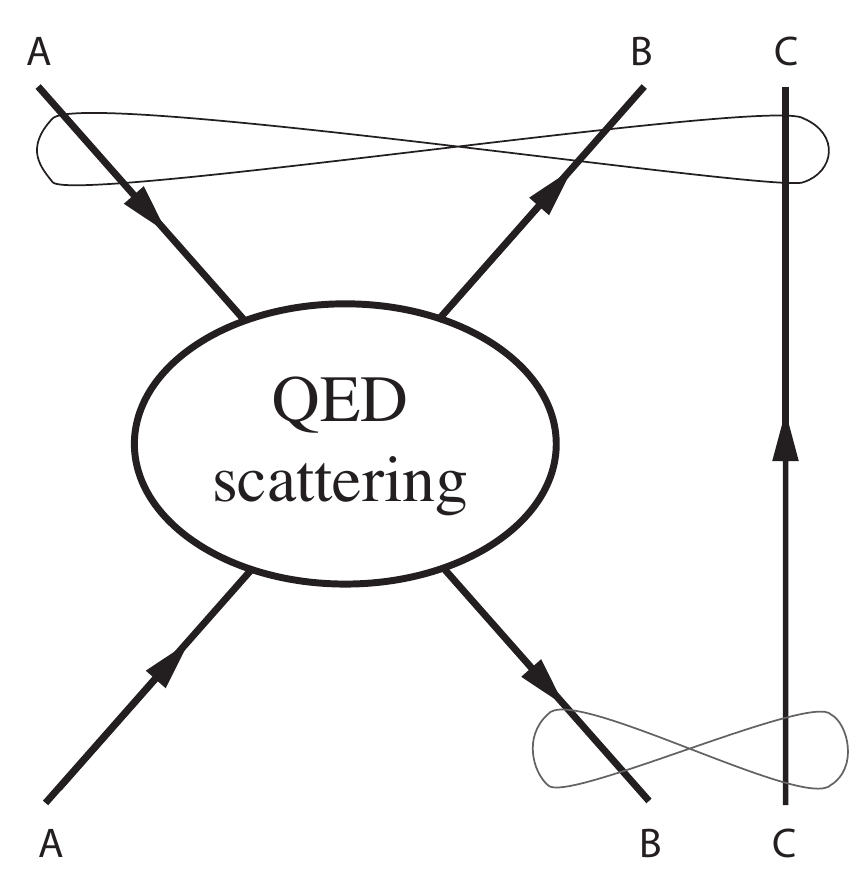}
\par\end{centering}
\caption{QED scattering with a witness particle. The particles $B$ and $C$ are initially entangled in spin. After the scattering, the three particles become entangled.}
\label{fig1}
\end{figure}

We consider a QED scattering involving $2$ particles, $A$ and $B$, in which $B$ is initially entangled in spin with $C$, i.e. the witness particle (see Fig. \ref{fig1}). The purpose is to evaluate the effect of the scattering over particle $C$, which does not take part directly in the scattering. We hope to extract information about the scattering by performing measurements on the subsystem $C$ after the process occurred.

In order to perform the calculations, a few definitions must be made. First, the internal product of fermion states is defined as
\begin{equation}
\langle k,a|p,b\rangle=2E_{\boldsymbol{k}}(2\pi)^{3}\delta^{(3)}(\boldsymbol{k}-\boldsymbol{p})\delta_{a,b},
\end{equation}
and, if a $2$-fermion system is initially in state
\begin{equation}
\vert \text{initial}\rangle=\vert{p}_{1},a;{p}_{2},b\rangle,
\end{equation}
after it undergoes a scattering process, it becomes
\begin{align}
\vert \text{final}\rangle=&\sum_{r,s}\int_{\boldsymbol{p}{}_{3},\boldsymbol{p}{}_{4}}\vert{p}_{3},r;{p}_{4},s\rangle \nonumber\\
&\times\langle{p}_{3},r;{p}_{4},s\vert S\vert{p}_{1},a;{p}_{2},b\rangle,
\label{fstatedef}
\end{align}
where the integral $\int_{\boldsymbol{p}}$ denotes $\int\left(d^{3}\boldsymbol{p}\right)/(2E_{\boldsymbol{p}}\left(2\pi\right)^{3})$. The $S$ matrix is written as $S=\mathcal{I}+i\mathcal{T}$, and the operator $i\mathcal{T}$ is related to the Feynman amplitude as
\begin{align}
\langle{p}_{3},r;{p}_{4},s\vert i\mathcal{T}\vert{p}_{1},a;&
{p}_{2},b\rangle=\,i\left(2\pi\right)^{4} \label{T} \\
&\times\delta^{\left(4\right)}\left({p}_{1}+{p}_{2}-{p}_{3}-{p}_{4}\right)\mathcal{M}_{i\rightarrow f}.\nonumber
\end{align}

We will consider an initial state as follows:
\begin{equation}
\vert i\rangle=\vert p_{1},a\rangle\otimes\left(\cos\eta\vert p_{2},\uparrow; q,\uparrow\rangle+\text{e}^{i\beta}\sin\eta\vert p_{2},\downarrow; q,\downarrow\rangle\right),
\label{instate}
\end{equation}
whose final state, according to Eqs. (\ref{fstatedef}) and (\ref{T}), is given by
\begin{align}
\vert &f\rangle =\vert i\rangle+i\sum_{r,s}\int_{\boldsymbol{p}{}_{3},\boldsymbol{p}{}_{4}\neq\boldsymbol{p}{}_{1},\boldsymbol{p}{}_{2}}\delta^{(4)}\left(p_1+p_2-p_3-p_4\right)\nonumber \\
&\times\Big[\cos\eta\mathcal{M}\left(a,\uparrow; r,s\right)\vert p{}_{3},r\rangle\otimes\vert p{}_{4},s\rangle\otimes\vert q,\uparrow\rangle \nonumber \\
 & +\text{e}^{i\beta}\sin\eta\mathcal{M}\left(a,\downarrow; r,s\right)\vert p{}_{3},r\rangle\otimes\vert p{}_{4},s\rangle\otimes\vert q,\downarrow\rangle\Big]
\label{FinalState}
\end{align}
where $\mathcal{M}\left(a,\uparrow; r,s\right)$ in fact denotes $\mathcal{M}
\left( p_{1},a;p_{2},\uparrow \mapsto p_{3},r;p_{4},s\right)$, but as all $\mathcal{M}$s have
the same dependence on initial and final momenta, these will be omitted for shortness, and we will leave only the spin dependence.

The final state of system $ABC$, but for a normalization constant ($\mathcal{N}$) yet to be calculated, is then
\begin{equation}
\rho_f=\vert f\rangle\langle f\vert.
\label{statef}
\end{equation}

As we are interested in $C$'s reduced density matrix, it is necessary to trace subsystems $A$ and $B$ out. The partial trace operation over a subsystem, say $b$, is illustrated below
\begin{equation}
\text{Tr}_{b}\left[\rho\right]=\sum_{\sigma}\int\frac{d^{3}\boldsymbol{k}}{(2\pi)^{3}}\frac{1}{2E_{\boldsymbol{k}}}\left(1_{r}\otimes\langle k,\sigma|_{b}\right)\rho\left(1_{r}\otimes|k,\sigma\rangle_{b}\right),
\end{equation}
where $1_r$ denotes the identity operation in the remaining subspaces. In performing partial traces, one finds Dirac deltas as $\left(2\pi \right)\delta^{(T)}\left(0\right)$ and $\left(2\pi\right)^3\delta^{\left(3\right)}\left(0\right)$, which enforce energy-momentum conservation. These have to be suitably regulated as described in Refs. \cite{deltas0,Seki2019}, using
\begin{align}
2\pi\delta^{(T)}\left(E_i-E_f\right)&=\int_{-T/2}^{T/2}\exp\left[{i\left(E_i-E_f\right)t}\right]dt \nonumber \\
\left(2\pi\right)^3\delta^{\left(3\right)}\left(\boldsymbol{k}-\boldsymbol{p}\right)&=V\delta_{\boldsymbol{k},\boldsymbol{p}},
\end{align}
which imply $\left(2\pi \right)\delta^{(T)}\left(0\right)=T$ and $\left(2\pi\right)^3\delta^{\left(3\right)}\left(0\right)=V$. Accordingly, the reduced density matrix 
of system $C$ is
\begin{equation}
\left(\rho_C\right)_f=\frac{\text{Tr}_A\left[\text{Tr}_B\left[\rho_f\right]\right]}{\mathcal{N}},
\end{equation}
where the numerator is
\begin{align}
\text{Tr}_A[\text{Tr}_B&[\rho_f]]=\Big[\left(2E_{\boldsymbol{p}_{1}}
2E_{\boldsymbol{p}_{2}}2E_{\boldsymbol{q}}V^{3}+2E_{\boldsymbol{q}}
TV^{2}\Lambda\right) \nonumber\\
&\times\left(\cos^{2}\eta\vert\uparrow\rangle\langle\uparrow\vert+\sin^{2}\eta\vert\downarrow\rangle\langle\downarrow\vert\right)\nonumber \\
 & +2E_{\boldsymbol{q}}TV^{2}\Lambda \cos\eta\sin\eta\nonumber\\
 &\times\left(\text{e}^{i\beta}\vert\downarrow\rangle\langle\uparrow\vert+\text{e}^{-i\beta}\vert\uparrow\rangle\langle\downarrow\vert\right)\Big]\otimes\frac{\vert q\rangle\langle q\vert}{2E_{\boldsymbol{q}}V},
\end{align}
and the factor $\Lambda$ (in fact $\eta$-dependent) reads
\begin{align}
\Lambda\left(\eta\right)=&\sum_{r,s}\int_{\boldsymbol{p}{}_{4}}\frac{T}{2E_{\boldsymbol{p}_{1}+\boldsymbol{p}_{2}-\boldsymbol{p}{}_{4}}}\big(\cos^{2}\eta\left|\mathcal{M}\left(a,\uparrow; r,s\right)\right|^{2}\nonumber\\
&+\sin^{2}\eta\left|\mathcal{M}\left(a,\downarrow; r,s\right)\right|^{2}\big)\big\vert_{\boldsymbol{p}_{3}=\boldsymbol{p}_{1}+\boldsymbol{p}_{2}-\boldsymbol{p}{}_{4}},\label{LambdaDef}
\end{align}
where we kept a factor of $T$ inside the integral, so as to perform the volume integrals in momentum space correctly. In the CM reference frame, one has $\int_{\boldsymbol{p}}T\equiv2\pi\delta\left(E_i-E_f\right)\times \left(2E_{\boldsymbol{p}}\right)^{2}d\Omega/\left(\left(2\pi\right)^{3}2E_{\boldsymbol{p}}\right)$, where $E_{\boldsymbol{p}}$ is the energy of any incoming/emerging particle.

The normalization is given by
\begin{align}
\mathcal{N}&=\text{Tr}_A\left[\text{Tr}_B\left[\text{Tr}_C\left[\left(\rho_{ABC}\right)_f\right]\right]\right]\nonumber\\
&=2E_{\boldsymbol{p}_{1}}2E_{\boldsymbol{p}_{2}}2E_{\boldsymbol{q}}V^{3}+2E_{\boldsymbol{q}}TV^{2}\Lambda,
\label{Normalization}
\end{align}
so as to ensure $\text{Tr}\left[\left(\rho_C\right)_f\right]=1$. Note that we factored the spin and momentum subspaces and wrote the momentum part as a projection operator, i.e. $\left[\vert q\rangle\langle q\vert/\left(2E_{\boldsymbol{q}}V\right)\right]^2=\vert q\rangle\langle q\vert/\left(2E_{\boldsymbol{q}}V\right)$. Below we investigate if it is possible read information about the scattering by measuring $C$.

\subsection{Inferring scattering data from the witness particle}
In order to extract information about the scattering from particle $C$, we begin by writing $\left(\rho_C\right)_f$ in matrix form
\begin{equation}
\left(\rho_{C}\right)_{f}=\begin{pmatrix}\cos^{2}\eta & \frac{e^{-i\beta}\Lambda T\sin\eta\cos\eta}{\Lambda T+2E_{\boldsymbol{p}_{1}}2E_{\boldsymbol{p}_{2}}V}\\
\frac{e^{i\beta}\Lambda T\sin\eta\cos\eta}{\Lambda T+2E_{\boldsymbol{p}_{1}}2E_{\boldsymbol{p}_{2}}V} & \sin^{2}\eta
\end{pmatrix},
\label{DMCfinal}
\end{equation}
from which we omitted the momentum subspace, $\vert q\rangle\langle q\vert/\left(2E_{\boldsymbol{q}}V\right)$. If compared to its initial density matrix, that is
\begin{equation}
\left(\rho_{C}\right)_{i}=\begin{pmatrix}\cos^{2}\eta & 0\\
0 & \sin^{2}\eta,
\end{pmatrix}
\end{equation}
it is evident that coherence (off-diagonal) terms were created in subsystem $C$, i.e. $C$ became purer. In addition, if one measures the initial and final expectation values of $\sigma_z$, one obtains
\begin{equation}
\langle \sigma_z\rangle_{i,f}=\cos^2\eta-\sin^2\eta.
\end{equation}

Regarding the initial expectation values of either $\sigma_x$ or $\sigma_y$, these are zero. However, if one performs these measures over the final state, one has
\begin{equation}
\langle \sigma_x\rangle_{f}=\cos\beta\sin\left(2\eta\right)\frac{\Lambda T}{\Lambda T+2E_{\boldsymbol{p}_1}2E_{\boldsymbol{p}_2}V},
\end{equation}
which, to first order in $\Lambda$, is
\begin{equation}
\langle\sigma_{x}\rangle_{f}=\cos\beta\sin\left(2\eta\right)\frac{\Lambda T}{{E_{CM}^2}V},
\label{expX}
\end{equation}
in the reference frame of the center of mass (CM), for which $E_{\boldsymbol{p}_1}=E_{\boldsymbol{p}_2}=E_{CM}^2/4$. From Eq. (\ref{expX}) we infer that $\langle\sigma_{x}\rangle_{f}$ is maximal if $B$ and $C$ are initially entangled as a Bell basis ($\eta=\pi/4$ and $\beta=0,\pi$). In other words, the choice of a Bell basis for $B$ and $C$ optimizes the effect of the $AB$ scattering over subsystem $C$.

\begin{figure}[h]
\begin{centering}
\includegraphics[scale=0.4]{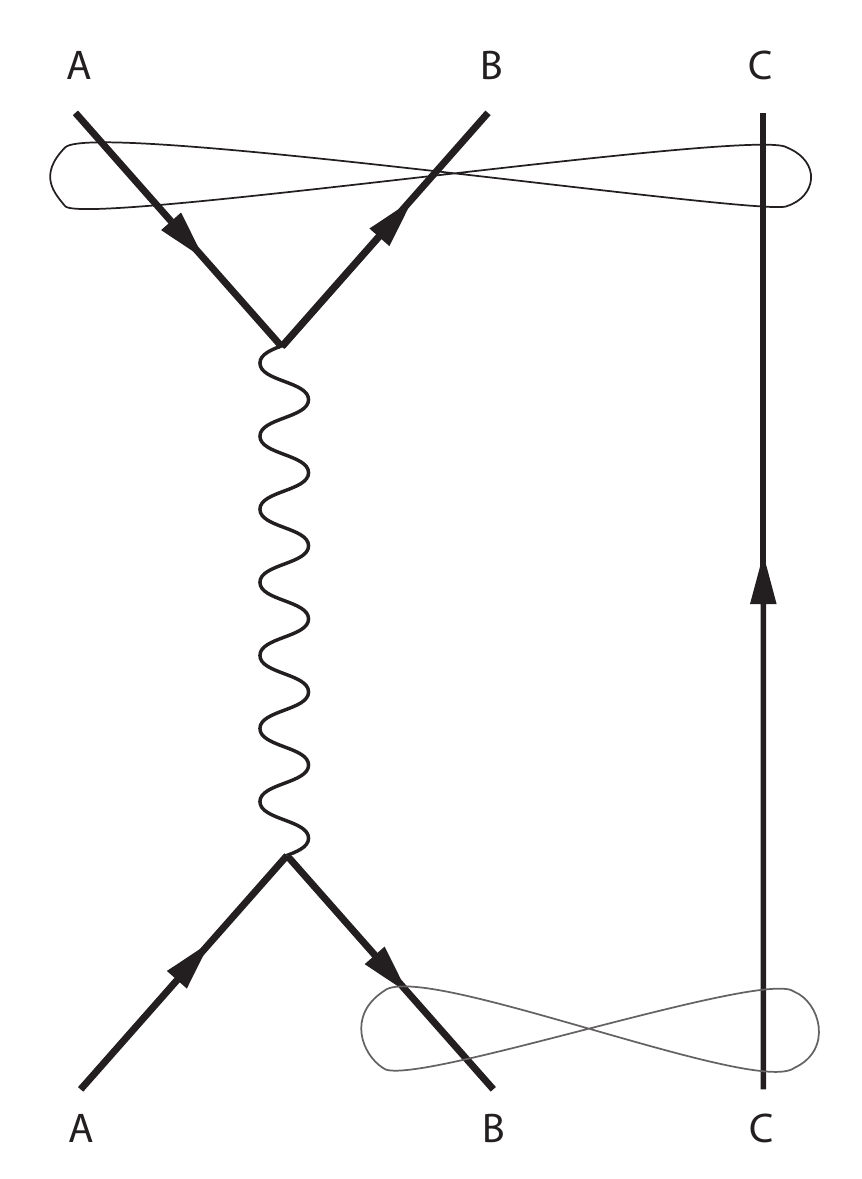}
\par\end{centering}
\caption{$e^+e^-\rightarrow\mu^+\mu^-$ scattering with a witness particle $C$. The quantity $\Lambda$ is found to be proportional to cross section if the $A$ beam is unpolarized.}
\label{fig2}
\end{figure}

Further, we could investigate the physical meaning of $\Lambda$. This is done by choosing a particular QED scattering and evaluating (\ref{LambdaDef}) at tree level. For this we consider the process $e^+e^-\rightarrow\mu^+\mu^-$ (see Fig. \ref{fig2}), in the CM reference frame.%
The momenta for the electron, positron, muon and anti-muon, are, respectively 
\begin{align}
p_1=&\left(E,0,0,p\right)\nonumber\\
p_2=&\left(E,0,0,-p\right)\nonumber\\
p_3=&\left(E,P\sin\theta\cos\phi,P\sin\theta\sin\phi,P\cos\theta\right)\nonumber\\
p_4=&\left(E,-P\sin\theta\cos\phi,-P\sin\theta\sin\phi,-P\cos\theta\right).
\label{momenta}
\end{align}

We found that, taking an unpolarized $A$ beam, which is equivalent to averaging over the spin $a$, the quantity $\Lambda$ is related to the total cross section ($\sigma_{\text{total}}$) of the process $e^+e^-\rightarrow\mu^+\mu^-$ as
\begin{equation}
\Lambda=\frac{4\vert\boldsymbol{p}_{1}\vert\left(E_{CM}\right)^{2}}{\vert\boldsymbol{p}_{3}\vert}\sigma_{\text{total}},
\end{equation}
so that, to first order, we have
\begin{equation}
\langle\sigma_{x}\rangle_{f}=4\sqrt{\frac{1-\frac{m^{2}}{E^{2}}}{1-\frac{M^{2}}{E^{2}}}}\sigma_{\text{total}}f\left(\eta,\beta\right),
\label{propCrossSection}
\end{equation}
where, $m$ ($M$) is the electron (muon) mass, $E$ is the energy of the incoming or emerging particles ($E=E_{CM}/2$), and $f\left(\eta,\beta\right)=\cos\beta\sin\left(2\eta\right)T/V$, which is maximal for a Bell basis.

The Eq. (\ref{propCrossSection}) implies that the spin of $C$ in the $x$-direction (or $y$-direction) is proportional to the total cross section of the scattering involving $A$ and $B$. Furthermore, Eq. (\ref{propCrossSection}) is valid in any energy regime for the scattering $e^+e^-\rightarrow\mu^+\mu^-$.

Next we investigate how the scattering modifies the distribution of mutual information among systems $A$, $B$ and $C$. From now on, for simplicity, we will adopt the ultrarelativistic limit in the CM reference frame, for which, $\sqrt{\left(1-\frac{m^{2}}{E^{2}}\right)/ \left(1-\frac{M^{2}}{E^{2}}\right)}\rightarrow1$ and $\Lambda\rightarrow e^4/(3\pi)$, and an unpolarized $A$ beam.

\section{Redistribution of mutual information}\label{Mutualinfo}
The mutual information is a quantifier of the total (quantum $+$ classical) correlations between two systems. Always non-negative, it is defined as
\begin{equation}
I_{XY}=S_X+S_Y-S_{XY},
\label{Defminfo}
\end{equation}
where $S_X$, $S_Y$, $S_{XY}$ stand for the von Neumann entropies of systems $X$, $Y$ and $XY$, respectively. It can be read as the amount of information that is contained in the system $XY$ that is not contained in the subsystems $X$ and $Y$, when taken separately; or what can one know about $X$ by measuring $Y$, and vice-versa. We choose to use this quantity for it is a more meaningful quantity when studying systems with ($N>2$)-parts. Another reason for doing so, is that the entropy of a fermion system, considering both spin and momentum of each particle, is Lorentz-invariant \cite{Bertlman2010}. We must point out that recently there have been advances in defining entanglement in $N>2$-partite systems \cite{Plastino2009,WitnessOp,Hassan2019}, and in continuous variable systems \cite{Duan}.
\subsection{Subsystem $AC$}
According to the definition of mutual information above, it is clear that the initial mutual information between $A$ and $C$ is zero - their subspaces are factored (see Eq. (\ref{instate})). Nevertheless, the final state entangles them via the initial entanglement between $B$ and $C$, implying that the mutual information between $A$ and $C$ should increase after the scattering. In order to verify this claim, it is necessary to evaluate the reduced density matrices of $A$ and $AC$, for we already have $C$'s final state in Eq. (\ref{DMCfinal}), and use definition (\ref{Defminfo}) to calculate the final mutual information between $A$ and $C$, $\left(I_{AC}\right)_f$. In evaluating $S_{AC}$, it is necessary obtain $\left(\rho_{AC}\right)_f$ by tracing out the system $B$ from the final state (\ref{statef}). After the partial trace over $B$, one has
\begin{equation}
\left(\rho_{AC}\right)_{f}=\frac{1}{\mathcal{N}}\left(I+II\right),
\label{rhoACf}
\end{equation}
where $I$ is
\begin{align}
I&=2E_{\boldsymbol{p}_{1}}2E_{\boldsymbol{p}_{2}}2E_{\boldsymbol{q}}V^{3}\Bigg[\frac{1}{2}\sum_{a}\vert a\rangle\langle a\vert\nonumber\\
&\otimes\left(\cos^{2}\eta\vert\uparrow\rangle\langle\uparrow\vert+\sin^{2}\eta\vert\downarrow\rangle\langle\downarrow\vert\right)\Bigg]\otimes\frac{\vert p_{1}\rangle\langle p_{1}\vert}{2E_{\boldsymbol{p}_{1}}V}\otimes\frac{\vert q\rangle\langle q\vert}{2E_{\boldsymbol{q}}V},
\end{align}
in which the $2\times2$ matrix in square brackets has eigenvalues
\begin{align}
g_1=\frac{\cos ^2\eta }{2}\ ,&\ \ \ g_2=\frac{\cos ^2\eta }{2}\nonumber\\
g_3=\frac{\sin ^2\eta }{2}\ ,&\ \ \ g_3=\frac{\sin ^2\eta }{2}.
\label{ACfree}
\end{align}

As for the term $II$, it reads
\begin{align}
II& =2E_{\boldsymbol{q}}V^{2}T\int_{\boldsymbol{p}{}_{3}}\frac{T}{2E_{\boldsymbol{p}_{1}+\boldsymbol{p}_{2}-\boldsymbol{p}_{3}}}\Bigg\{\frac{1}{2}\sum_{a,s,r,r'}\nonumber \\
 & \Big[\cos^{2}\eta\mathcal{M}\left(a,\uparrow; r,s\right)\mathcal{M}^{*}\left(a,\uparrow; r',s\right)\vert r\rangle\langle r'\vert\otimes\vert\uparrow\rangle\langle\uparrow\vert\nonumber \\
 & +\text{e}^{-i\beta}\cos\eta\sin\eta\mathcal{M}\left(a,\uparrow; r,s\right)\mathcal{M}^{*}\left(a,\downarrow; r',s\right)\vert r\rangle\langle r'\vert\otimes\vert\uparrow\rangle\langle\downarrow\vert\nonumber \\
 & +\text{e}^{i\beta}\cos\eta\sin\eta\mathcal{M}\left(a,\downarrow; r,s\right)\mathcal{M}^{*}\left(a,\uparrow; r',s\right)\vert r\rangle\langle r'\vert\otimes\vert\downarrow\rangle\langle\uparrow\vert\nonumber \\
 & +\sin^{2}\eta\mathcal{M}\left(a,\downarrow; r,s\right)\mathcal{M}^{*}\left(a,\downarrow; r',s\right)\vert r\rangle\langle r'\vert\otimes\vert\downarrow\rangle\langle\downarrow\vert\Big]\Bigg\}\nonumber\\
& \otimes\frac{\vert p{}_{3}\rangle\langle p{}_{3}\vert}{2E_{\boldsymbol{p}_{3}}V}\otimes\frac{\vert q\rangle\langle q\vert}{2E_{\boldsymbol{q}}}.
 \label{trBinteracting}
\end{align}

In the ultrarelativistic limit ($m,M\rightarrow0$ and $p,P\rightarrow E$ in Eq. (\ref{momenta})), the eigenvalues of the $4\times4$ matrix in curly brackets of Eq. (\ref{trBinteracting}) are
\begin{align}
\mathcal{M}_{AC1} & =2e^{4}\cos^{2}\eta\cos^{4}\left(\frac{\theta}{2}\right)\nonumber\\
\mathcal{M}_{AC2} & =2e^{4}\sin^{2}\eta\cos^{4}\left(\frac{\theta}{2}\right)\nonumber\\
\mathcal{M}_{AC3} & =2e^{4}\cos^{2}\eta\sin^{4}\left(\frac{\theta}{2}\right)\nonumber\\
\mathcal{M}_{AC4} & =2e^{4}\sin^{2}\eta\sin^{4}\left(\frac{\theta}{2}\right).
\label{ACint}
\end{align}

Using the normalization (\ref{Normalization}) and the eigenvalues (\ref{ACfree},\ref{ACint}), one can calculate $AC$'s final entropy as
\begin{equation}
\left(S_{AC}\right)_f=-\sum_{i}^{4} \left[G_i \ln G_i+\int d\Omega \left( \tilde{\mathcal{M}}_{ACi} \ln \tilde{\mathcal{M}}_{ACi}\right)\right],
\label{SACf}
\end{equation}
where
\begin{equation}
G_i=\frac{g_i}{1+\frac{T}{V}\frac{\Lambda}{4E^2}},
\end{equation}
and
\begin{equation}
\tilde{\mathcal{M}}_{ACi}=\left(\frac{1}{\frac{4E^2V}{T}+\Lambda}\right)\frac{\mathcal{M}_{ACi}}{4\left(2\pi\right)^2}
\end{equation}
were calculated in the center of mass in the ultrarelativistic limit, for which $\Lambda=e^4/(3\pi)$. We calculate next the final reduced density matrix of particle $A$, which is done by tracing particle $C$ out of (\ref{rhoACf}), yielding
\begin{align}
\left(\rho_A\right)_f=\frac{1}{\mathcal{N}}\left(III+ IV\right),
\end{align}
where
\begin{align}
III=2E_{\boldsymbol{p}_{1}}2E_{\boldsymbol{p}_{2}}2E_{\boldsymbol{q}}V^{3}\left[\frac{1}{2}\sum_{a}\vert a\rangle\langle a\vert\right]\otimes\frac{\vert p_1\rangle\langle p_1\vert}{2E_{\boldsymbol{p}_{1}}V},
\end{align}
is already diagonal, and
\begin{align}
&IV =2E_{\boldsymbol{q}}V^2T\int_{\boldsymbol{p}_{3}}\frac{T}{2E_{\boldsymbol{p}_{1}+\boldsymbol{p}_{2}-\boldsymbol{p}_{3}}}\Bigg[\frac{1}
{2} \sum_{a,s,r,r'} \nonumber \\
 & \Big(\cos^{2}\eta\mathcal{M}\left(a,\uparrow; r,s\right)\mathcal{M}^{*}\left(a,\uparrow; r',s\right)\vert r\rangle\langle r'\vert\nonumber \\
 & +\sin^{2}\eta\mathcal{M}^{*}\left(a,\downarrow; r',s\right)\mathcal{M}\left(a,\downarrow; r,s\right)\vert r\rangle\langle r'\vert\Big) \Bigg]\otimes\frac{\vert p_3\rangle \langle p_3\vert}{2E_{\boldsymbol{p}_{3}}V},
 \label{Aint}
\end{align}
needs to be diagonalized. The eigenvalues of the matrix in square brackets in Eq. (\ref{Aint}) are
\begin{align}
\mathcal{M}_{A1} & =\frac{1}{4}e^{4}\left(\cos2\theta+3+4\cos2\eta\cos\theta\right)\nonumber\\
\mathcal{M}_{A2} & =\frac{1}{4}e^{4}\left(\cos2\theta+3-4\cos2\eta\cos\theta\right).
\label{EigsAint}
\end{align}

The final entropy of $A$ is then
\begin{equation}
\left(S_A\right)_f=-2h\ln h-\sum_{i}^{2}\int d\Omega \left( \tilde{\mathcal{M}}_{Ai} \ln \tilde{\mathcal{M}}_{Ai}\right),
\label{SAf}
\end{equation}
in which
\begin{equation}
h=\frac{1}{2}\left(\frac{1}{1+\frac{T}{V}\frac{\Lambda}{4E^2}}\right),
\end{equation}
and
\begin{equation}
\tilde{\mathcal{M}}_{Ai}=\left(\frac{1}{\frac{4E^2V}{T}+\Lambda}\right)\frac{\mathcal{M}_{Ai}}{4\left(2\pi\right)^2}\ .
\end{equation}

The final entropy of system $C$ reads
\begin{equation}
\left(S_C\right)_f=-\sum_{i}^{2}c_i\ln c_i\ ,
\label{SCf}
\end{equation}
in which $c_i$ correspond to the eigenvalues of the density matrix (\ref{DMCfinal}). The final
 mutual information between $A$ and $C$, using Eqs. (\ref{SACf}), (\ref{SAf}) and 
 (\ref{SCf}), is
\begin{equation}
\left(I_{AC}\right)_f=\left(S_A\right)_f+\left(S_C\right)_f-\left(S_{AC}\right)_f.
\label{IACf}
\end{equation}

\begin{figure}[h]
\begin{centering}
\includegraphics[scale=0.7]{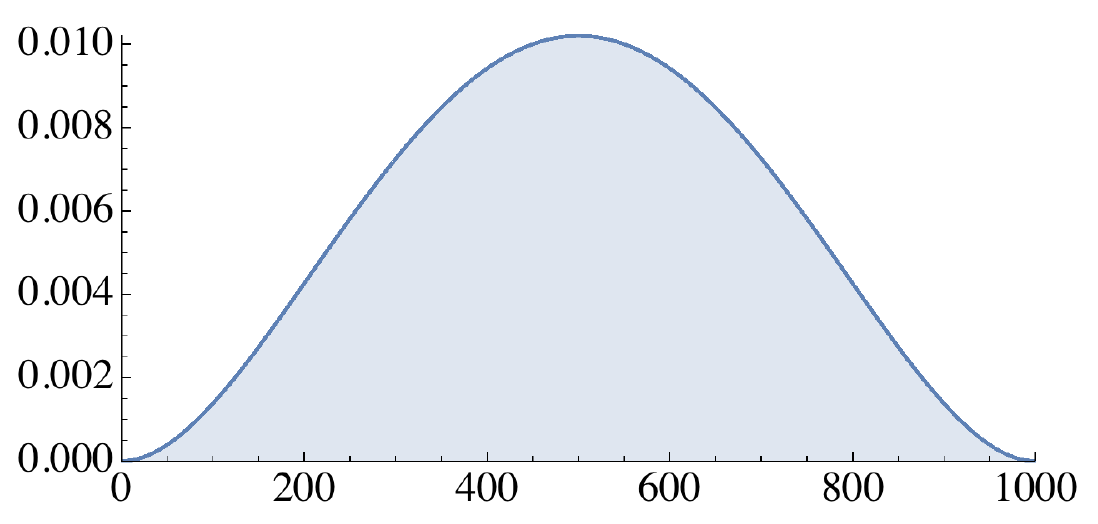}
\par\end{centering}
\caption{Plot of $\left(I_{AC}\right)_f$ for the set of parameters ($T,\,V,\,E_{\boldsymbol{p}_{i}},\,e\rightarrow1;\,\Lambda\rightarrow e^{4}/3\pi$). The angle $\eta$ was split in $n=1000$ parts from $0$ to $\pi/2$ in order to run the plot. The final mutual information is maximal for $\eta=\pi/4$ -- compatible with a Bell basis for $BC$'s initial state. Conversely, for $\eta=0$ or $\eta=\pi/2$, i.e. $B$ and $C$ initially unentangled, there is no mutual information between $A$ and $C$.}
\label{plotIACf}
\end{figure}

A plot of  (\ref{IACf}) is shown in Fig. \ref{plotIACf}.
Below we perform this analysis over another partition of the system.
\subsection{Subsystem $BC$}
It would be interesting to evaluate how the mutual information varies in other partitions of the system, such as $BC$. Unlike partition $AC$, $B$ and $C$ are initially entangled, so that their initial mutual information is not zero. Using state (\ref{instate}), one obtains
\begin{align}
\left(I_{BC}\right)&_i=\left(S_{B}\right)_i+\left(S_{C}\right)_i\nonumber\\
&=-2\left[\cos^2\eta\ln\left( \cos^2\eta\right) +\sin^2\eta\ln \left(\sin^2\eta\right)\right],
\end{align}
where we omitted $\left(S_{BC}\right)_i$ for it is null. By tracing $A$ out of the final state, we obtain $BC$'s reduced density matrix, that is
\begin{equation}
\left(\rho_{BC}\right)_{f}=\frac{1}{\mathcal{N}}\left(\overline{I}+\overline{II}\right),
\label{rhoBCf}
\end{equation}
 in which
\begin{align}
\overline{I}&=2E_{\boldsymbol{p}_{1}}2E_{\boldsymbol{p}_{2}}2E_{\boldsymbol{q}}V^{3}\Bigg[\cos^{2}\eta\vert\uparrow\rangle\langle\uparrow\vert\otimes\vert\uparrow\rangle\langle\uparrow\vert\nonumber\\
&+\frac{1}{2}\sin2\eta\Big(\text{e}^{-i\beta}\vert\uparrow\rangle\langle\downarrow\vert\otimes\vert\uparrow\rangle\langle\downarrow\vert+\text{e}^{i\beta}\vert\downarrow\rangle\langle\uparrow\vert\otimes\vert\downarrow\rangle\langle\uparrow\vert\Big)\nonumber\\
&\sin^{2}\eta\vert\downarrow\rangle\langle\downarrow\vert\otimes\vert\downarrow\rangle\langle\downarrow\vert\Bigg]\otimes\frac{\vert p_{2}\rangle\langle p_{2}\vert}{2E_{\boldsymbol{p}_{2}}V}\otimes\frac{\vert q\rangle\langle q\vert}{2E_{\boldsymbol{q}}V},
\label{BCfree}
\end{align}
and
\begin{align}
\overline{II}&=2E_{\boldsymbol{q}}V^{2}T
\int_{\boldsymbol{p}{}_{4}}\frac{T}{2E_{\boldsymbol{p}_{1}+\boldsymbol{p}_{2}-\boldsymbol{p}{}_{4}}}\Bigg\{\frac{1}{2}\sum_{a,r,s,s'}\nonumber \\
 & \Bigg(\cos^{2}\eta\mathcal{M}\left(a,\uparrow; r,s\right)\mathcal{M}^{*}\left(a,\uparrow; r,s'\right)\vert s\rangle\langle s'\vert\otimes\vert\uparrow\rangle\langle\uparrow\vert\nonumber \\
 & +\frac{\text{e}^{-i\beta}\sin2\eta}{2}\mathcal{M}\left(a,\uparrow; r,s\right)\mathcal{M}^{*}\left(a,\downarrow; r,s'\right)\vert s\rangle\langle s'\vert\otimes\vert\uparrow\rangle\langle\downarrow\vert\nonumber \\
 & +\frac{\text{e}^{i\beta}\sin2\eta}{2}\mathcal{M}\left(a,\downarrow; r,s\right)\mathcal{M}^{*}\left(a,\uparrow; r,s'\right)\vert s\rangle\langle s'\vert\otimes\vert\downarrow\rangle\langle\uparrow\vert\nonumber \\
 & +\sin^{2}\eta\mathcal{M}\left(a,\downarrow; r,s\right)\mathcal{M}^{*}\left(a,\downarrow; r,s'\right)\vert s\rangle\langle s'\vert\otimes\vert\downarrow\rangle\langle\downarrow\vert\Bigg)\Bigg\}\nonumber\\
 &\otimes\frac{\vert p_{4}\rangle\langle p_{4}\vert}{2E_{\boldsymbol{p}_{4}}V}\otimes\frac{\vert q\rangle\langle q\vert}{2E_{\boldsymbol{q}}V}.
 \label{BCint}
\end{align}

The matrix in square brackets in Eq. (\ref{BCfree}) has eigenvalues $\{0,0,0,1\}$, while the one in curly brackets in Eq. (\ref{BCint}) has the eigenvalues already listed in (\ref{ACint}). As for the final reduced density matrix of $B$, we have
\begin{align}
\left(\rho_B\right)_f=\frac{1}{\mathcal{N}}\left(\overline{III}+ \overline{IV}\right),
\label{rhoBf}
\end{align}
where
\begin{align}
\overline{III}&=2E_{\boldsymbol{p}_{1}}2E_{\boldsymbol{p}_{2}}2E_{\boldsymbol{q}}V^{3}\nonumber\\
&\times\left(\cos^{2}\eta\vert \uparrow\rangle\langle \uparrow\vert+\sin^{2}\eta\vert \downarrow\rangle\langle\downarrow\vert\right)\otimes\frac{\vert p_2\rangle\langle p_2\vert}{2E_{\boldsymbol{p}_{2}}V},
\label{Bfree}
\end{align}
 and
\begin{align}
&\overline{IV} =2E_{\boldsymbol{q}}V^2T\int_{\boldsymbol{p}_{4}}\frac{T}{2E_{\boldsymbol{p}_{1}+\boldsymbol{p}_{2}-\boldsymbol{p}_{4}}}\Bigg[\frac{1}
{2} \sum_{a,r,s,s'} \nonumber \\
 & \Big(\cos^{2}\eta\mathcal{M}\left(a,\uparrow; r,s\right)\mathcal{M}^{*}\left(a,\uparrow; r,s'\right)\vert s\rangle\langle s'\vert\nonumber \\
 & +\sin^{2}\eta\mathcal{M}^{*}\left(a,\downarrow; r,s'\right)\mathcal{M}\left(a,\downarrow; r,s\right)\vert s\rangle\langle s'\vert\Big) \Bigg]\otimes\frac{\vert p_4\rangle \langle p_4\vert}{2E_{\boldsymbol{p}_{4}}V}.
 \label{Bint}
 \end{align}

\begin{figure}[h]
\begin{centering}
\includegraphics[scale=0.6]{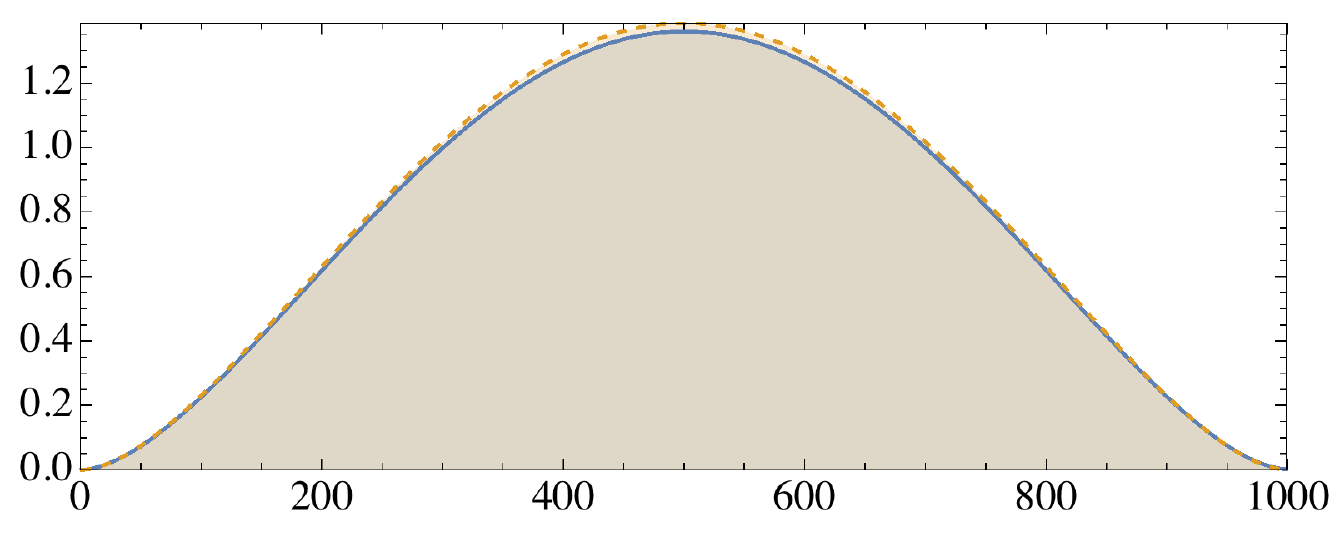}
\par\end{centering}
\caption{Initial (dashed line) and final (thick line) mutual Information between
$B$ and $C$. There is a decrease, which is largest for $\eta=\pi/4$, for part of the
correlations are transferred to the partition $AC$. The plot was made using the set of parameters ($T,\,V,\,E_{\boldsymbol{p}_{i}},\,e\rightarrow1;\,\Lambda\rightarrow e^{4}/3\pi$). The angle $\eta$ was split in $n=1000$ parts from $0$ to $\pi/2$ in order to run the plot.}
\label{IBCif}
\end{figure}

We can now use the  final density matrices of the subsystem $BC$, in (\ref{DMCfinal},\ref{rhoBf},\ref{rhoBCf}) to calculate the mutual information between $C$ and $C$ after the scattering. A plot of the initial and final mutual information $I_{BC}$ is shown in Fig. \ref{IBCif}.

\section{Conclusions and final remarks}\label{Conclusions}

We analyzed a QED scattering $AB\rightarrow AB$, in which $B$ was initially entangled with a third particle $C$ that did not participate directly in process. After calculating the reduced density matrix of particle $C$, we found that coherence (off-diagonal) terms were created. Although these do not change its spin expectation value in the
 $z$-direction, in orthogonal directions we obtain, for instance, $\langle\sigma_{x}\rangle_{f}\propto\sigma_{\text{total}}f\left(\eta,\beta\right)$,
 in which $f\left(\eta,\beta\right)$ is maximal for $BC$ initially entangled as a Bell basis. We
 point out that the factor $\Lambda$ is $\eta$-independent only if we consider an initially 
 unpolarized $A$ beam. That said, the result indicate that, at least in principle, one could 
 measure the total cross section of scattering $AB\rightarrow AB$  letting particle $C$ go through a Stern-Gerlach apparatus. This method could be used to measure cross 
 sections when the products $A$ and/or $B$ are cumbersome to detect.
 
Next we studied the effect of the scattering on the amount of correlations between different partitions of the system. Initially the system is entangled only in the subspace spanned by $BC$; after the scattering, all three subsystems are entangled. In order to describe the correlation transfer, we chose to calculate the mutual information between $A$ and $C$, and between $B$ and $C$. This quantity, being written in terms of von Neumann entropies of the subsystem formed the particles' momenta and spins, taken together, is Lorentz-invariant.

We found that there is an increase (decrease) in the mutual information between $A$ and $C$ ($B$ and $C$) which is largest for $\eta=\pi/4$ -- compatible with a Bell basis. The largest decrease in the mutual information between $B$ and $C$ is of about $2\%$ for the set of parameters chosen ($T,\,V,\,E_{\boldsymbol{p}_{i}},\,e\rightarrow1;\,\Lambda\rightarrow e^{4}/3\pi$). This decrease in $I_{BC}$ does not match, however, the increase in $I_{AC}$, for after the scattering there will be mutual information between $A$ and $B$. In addition, unlike the expectation value $\langle\sigma_{x}\rangle_{f}$, these quantities are dependent only on the mixing angle $\eta$ -- they are not sensitive to the phase $\beta$.

\section{Acknowledgements}

The authors are thankful to Capes, CNPq, CFisUC and Fundação para Ciência e Tecnologia (FCT) (through the project UID/FIS/04564/2016) for the financial support.

\end{document}